\documentclass[pdflatex,sn-basic]{sn-jnl} 

\jyear{2025}%

\theoremstyle{thmstyleone}

\theoremstyle{thmstyletwo}

\theoremstyle{thmstylethree}

\usepackage{amssymb}
\usepackage{pifont}

\usepackage{amsmath}
\usepackage{lineno}
\usepackage{accsupp}
\usepackage{fancyhdr}
\usepackage{lastpage}
\usepackage{adjustbox}
\usepackage{tabularx}
\usepackage{makecell}
\usepackage{multirow}
\usepackage[T1]{fontenc}
\usepackage{lmodern}
\usepackage{anyfontsize}

\begin{document}

\title[Article Title]{Pluri-Gaussian rapid updating of geological domains}

\author*[1,2]{\fnm{Sultan} \sur{Abulkhair}}\email{sultan.abulkhair@adelaide.edu.au}

\author[1,2]{\fnm{Peter} \sur{Dowd}}\nomail

\author[2]{\fnm{Chaoshui} \sur{Xu}}\nomail

\affil[1]{\orgdiv{ARC Training Centre for Integrated Operations for Complex Resources}, \orgname{The University of Adelaide}, \orgaddress{\city{Adelaide, SA 5005}, \country{Australia}}}

\affil[2]{\orgdiv{School of Chemical Engineering}, \orgname{The University of Adelaide}, \orgaddress{\city{Adelaide, SA 5005}, \country{Australia}}}

\abstract{Over the past decade, the rapid updating of resource knowledge and the integration of real-time sensor information have gained attention in both industry and academia. However, most studies on rapid resource model updating have focused on continuous variables, such as grade variables and coal quality parameters. Geological domain modelling is an essential component of resource estimation, which is why it is crucial to extend data assimilation techniques to enable the rapid updating of categorical variables. In this paper, a methodology inspired by pluri-Gaussian simulation is proposed for near-real-time updating of geological domains, followed by updating grade variables within these domain boundaries. The proposed algorithm consists of a Gibbs sampler for converting geological domains into Gaussian random fields, an ensemble Kalman filter with multiple data assimilations for rapid updating, and rotation based iterative Gaussianisation for multi-Gaussian transformation. We demonstrate the algorithm by using a synthetic case study with observations sampled from the ground truth, as well as a real case study that uses production drilling samples to jointly update geological domains and grade variables. Both case studies are based on real data from an iron oxide-copper-gold deposit in South Australia. This approach enhances resource knowledge by incorporating both categorical and continuous variables, leading to improved reproduction of domain geometries, closer matches between predictions and observations, and more geologically realistic resource models.}


\maketitle

\section{Introduction}\label{sec1}

Resource modelling is a fundamental aspect of mining as it describes the location, quality, and quantity of mineral resources, which are essential for mine planning and optimisation \citep{bib1}. An important part of resource modelling is segmenting the ore body into spatial domains that have distinct geological and statistical characteristics. One way to account for these different domains and their grade distributions is through cascade modelling. In this method, the domains are modelled first, followed by estimating or simulating the grades within each domain \citep{bib2,bib3,bib4,bib5,bib6}.

Conventional deterministic methods, such as wireframing, can produce accurate models when the drill hole spacing is dense and the geological domains are not overly complex \citep{bib2,bib7}. However, in practice, mining deposits often involve multiple interconnected domains, and the available data may not be sufficiently dense to build precise geological models. In contrast, stochastic geostatistical methods can generate multiple equally probable realisations of these domains while also quantifying uncertainty, particularly around the domain boundaries. Techniques such as sequential indicator simulation (SIS) \citep{bib8,bib9} and pluri-Gaussian simulation (PGS) \citep{bib10,bib11,bib12} are commonly used for this purpose.

PGS is considered to be a superior method compared to SIS due to its capability to incorporate contact relationships across multiple categories. This is achieved by representing geological domains as two or more Gaussian random fields (GRFs), using techniques such as Gibbs sampling to enforce these contact relationships \citep{bib11}. \cite{bib13} introduced an improved hierarchical approach that can use more than two GRFs and local proportions governed by an interpreted geological model. Additionally, truncating non-stationary GRFs enhances the effectiveness of PGS in complex and heterogeneous environments, such as porphyry copper deposits \citep{bib14}.

Another alternative is multiple-point statistics (MPS) methods, which provide better connectivity reproduction, especially for heterogeneous and long-range structures \citep{bib15}. For instance, the direct sampling (DeeSse) algorithm applies a distance function to directly scan the training image and obtain spatial variability information \citep{bib16}. Aside from its applications in hydrogeology and petroleum facies modelling, DeeSse has been applied to model complex domains such as barren dykes \citep{bib17}, coal seam layers \citep{bib5} and gold veins \citep{bib6}. The latter study showed considerably more accurate modelling of thin, long-range veins compared to PGS. However, the crucial downside of MPS remains its high simulation runtime and difficulty in deriving an appropriate training image, which should be large enough and sufficiently representative to generate accurate realisations.

Nevertheless, regardless of the choice of modelling methodology, geological uncertainty remains a significant challenge, particularly around domain boundaries. Any of the aforementioned methods can only ensure precisely accurate results at the drill hole locations. As mining operations go deeper and grades decrease, it becomes increasingly important to ensure accurate resource modelling. One effective way to achieve this is by integrating real-time sensor observations that are available during mining operations at various stages of the mining value chain. For example, the ensemble Kalman filter (EnKF) has been extensively applied in the rapid updating of resource and grade control models over the past decade \citep{bib18,bib19,bib20,bib21,bib22,bib23}. Alternatively, methods, such as residual kriging to update GRF realisations \citep{bib24}, actor-critic reinforcement learning \citep{bib25}, and EnKF with multiple data assimilations (EnKF-MDA) \citep{bib26} have also been proposed for the same task. For instance, EnKF-MDA updates the same data multiple times with an inflated measurement error \citep{bib27,bib28}. It offers a significantly closer match between model-based predictions and observations, especially in highly heterogeneous cases.

Most studies of rapid resource model updating, however, have focused on continuous variables (e.g., grades, coal quality parameters, and compositional data). The main challenge with categorical variables (e.g., lithology, alterations, and facies) is the complexity in transforming them into a format suitable for data assimilation with EnKF. Due to the Gaussianity assumption of EnKF, geological domains have to be parameterised into a continuous multivariate Gaussian space. For instance, \cite{bib23} proposed a discrete wavelet transform to turn indicator variables into frequency coefficients, followed by an EnKF rapid updating. However, it should be noted that the results validating this approach involved only two domains, making its applicability to cases with complex contact relationships across multiple domains unclear. Similarly, several methods have been proposed for the history matching of facies models, including PGS \citep{bib29}, principal component analysis (PCA) \citep{bib30}, level set functions \citep{bib31}, and deep generative models \citep{bib32}.

This study continues the work of \cite{bib26}, who proposed an algorithm that combines EnKF-MDA with a multivariate Gaussian transformation to update multivariate resource models rapidly. EnKF-MDA minimised deviations between predictions and observations compared to conventional EnKF, while rotation based iterative Gaussianisation (RBIG) \citep{bib33,bib34,bib35} helped to take complex multivariate relationships into account. In this paper, we extend this algorithm to update geological domains using a methodology inspired by PGS, in which new categorical observations are parameterised using a Gibbs sampler to rapidly update prior GRFs, followed by truncation. Cross-correlated grades are then updated within each domain, allowing for joint updating and the reduction of the uncertainty associated with domain boundaries and corresponding grade distributions.

The next section presents a detailed methodology for PGS, rapid multivariate resource-model updating using EnKF-MDA and RBIG, and the proposed pluri-Gaussian rapid updating algorithm. Results include both synthetic and real case studies based on real data from an iron oxide-copper-gold (IOCG) deposit in South Australia. In a synthetic study, 2D prior PGS realisations are updated using observations sampled from the ground truth model. On the other hand, real reverse circulation (RC) drilling samples, containing geological domains and three grade variables, are used as observations in a real case study. Finally, the paper concludes with a summary of results, potential limitations and future research directions.

\section{Methodology}\label{sec2}

\subsection{Pluri-Gaussian simulation}\label{subsec2.1}

To simplify the rapid updating process, it is assumed that prior realisations of geological domains are modelled using PGS. In this way, the proposed data assimilation algorithm will have the simulations of GRFs to work with as prior model-based predictions. The brief workflow of PGS used for this study is as follows:

\begin{enumerate}[1.]

\item Defining the truncation rule, which maps the contact relationships between domains. Based on the proportions of each domain, the truncation thresholds (i.e., the upper and lower bounds of GRFs for that specific domain) can be derived. In this paper, the number of GRFs is limited to two, and thus the truncation rule is defined by the 2D Gaussian space.

\item Modelling variograms of GRFs by either inferring them from the mathematical relationship between indicator variograms and variograms of standard GRFs \citep{bib11} or directly estimating them empirically using pairways likelihood \citep{bib36}.

\item Converting indicator variables of the conditioning data into GRFs via the Gibbs sampler \citep{bib44} using the proportion thresholds from the truncation rule and variogram models of GRFs. This process can be defined as follows for a particular domain and is repeated for a set number of iterations:
\begin{equation}
\Upsilon(x)=\Upsilon_K(x)+\sigma_KR(x),\label{eq1}
\end{equation}
where $\Upsilon_K(x)$ and $\sigma_K$ are respectively, the kriging estimate and its standard deviation at the selected data point conditioned to other data points, $R(x)$ is a residual obtained from a standard normal distribution $N(0,1)$ based on $\left[\frac{a_{min}-\Upsilon_K(x)}{\sigma_K},\frac{a_{max}-\Upsilon_K(x)}{\sigma_K}\right]$, and $[a_{min},a_{max}]$ is the interval from the truncation thresholds of a domain.

\item Simulating the GRFs at the block model locations, conditional to the Gaussian values obtained from step 3. This can be achieved using geostatistical simulation methods, such as sequential Gaussian simulation and turning bands simulation \citep{bib37,bib38}.

\item Truncating the simulated realisations of GRFs to obtain the realisations of domains according to the truncation rule defined in step 1.

\end{enumerate}

\subsection{EnKF-MDA and RBIG for rapid updating}\label{subsec2.2}

There are many viable options for the rapid updating of continuous variables. For instance, EnKF-MDA paired with a multi-Gaussian transformation offers improved data assimilation accuracy for updating multivariate resource models \citep{bib26}. Several algorithms can be used for the multi-Gaussian transformation, including the projection pursuit multivariate transform (PPMT) \citep{bib39}, flow transformation (FA) \citep{bib40}, and RBIG \citep{bib33}. RBIG is considerably faster than the others and achieves a performance comparable to PPMT and FA, making it an optimal method for time-sensitive tasks such as rapid updating \citep{bib34}. The steps of the combined EnKF-MDA and the RBIG algorithm are:

\begin{enumerate}[1.]

\item Extracting a block model neighbourhood around the available observations, defined by the pre-selected number of blocks from each observation.

\item Transforming a combined vector of n neighbourhood realisations and observations with m number of variables at time t into independent multi-Gaussian factors $\left({Z_{m}^{n,t}}^{\text{RBIG}},{O_{m}^{t}}^{\text{RBIG}}\right)$ using RBIG and storing all the iterations of transformation functions and rotation matrices as $F_{\text{RBIG}}^{t}$. The number of iterations depends on the complexity of the multivariate relationships, which may require more iterations to achieve convergence in complex cases. Additionally, stopping criteria can be added to avoid unnecessary computation. The following is a single iteration of RBIG:
\begin{equation}
\left(Z_{m}^{n,i+1},O_{m}^{i+1}\right)=R^{i}\Phi^{i}\left(Z_{m}^{n,i},O_{m}^{i}\right),\label{eq2}
\end{equation}
where $\Phi^{i}$ is a marginal Gaussianisation based on histogram equalisation at iteration $i$, $R^{i}$ is a PCA rotation matrix, $Z_{m}^{n,i}$ are $n$ neighbourhood realisations, and $O_{m}^{i}$ are observations with a number of variables $m$.

\item Computing the model-based predictions $H_{m}^{n,t}$ at observation locations.

\item Adding a measurement error $\varepsilon$ to observations.

\item Calculating the Gaspari-Cohn correlation filter for covariance localisation
\begin{equation}
\alpha(r)=
\begin{cases}
-\frac{1}{4}r^{5}+\frac{1}{2}r^{4}+\frac{5}{8}r^{3}-\frac{5}{2}r^{2}+1 & 0\le r<1\\
\frac{1}{12}r^{5}-\frac{1}{2}r^{4}+\frac{5}{8}r^{3}+\frac{5}{3}r^{2}-5r+4-\frac{2}{3}r^{-1} & 1<r\le2\\
0 & \hphantom{.1<}r>2
\end{cases},\label{eq3}
\end{equation}
where $r$ is a normalised distance between two locations defined by $\frac{d}{L}$, $d$ is a distance and $L$ is a predefined localisation radius. The localisation radius is problem-specific, and selecting the optimal one depends on the observation density, ensemble size, and model dimensions. The practical approach is to empirically tune it after trying different radius values.

\item Compute the Kalman gain
\begin{equation}
K_{m}^{t}=\alpha(r)C_{YD}^{m,t}\left(\alpha(r)C_{DD}^{m,t}+C_{D}^{m,t}\right)^{-1},\label{eq4}
\end{equation}
where $C_{YD}^{m,t}$ is the experimental covariance between realisations and model-based predictions, $C_{DD}^{m,t}$ is the experimental covariance of model-based predictions and $C_{D}^{m,t}$ is the experimental covariance of observations.

\item Update the prior realisations
\begin{equation}
{Z_{m}^{n,t+1}}^\text{RBIG}={Z_{m}^{n,t}}^\text{RBIG}+K_{m}^{t}\left({O_{m}^{t}}^{\text{RBIG}}-H_{m}^{n,t}\right).\label{5}
\end{equation}

\item Repeating steps 3-7 for a predefined number of data assimilations and ensuring that the random error added to observations differs from previous data assimilations. Generally, 4-5 data assimilations provide a good balance between accuracy and computation time.

\item Back-transforming updated realisations into the original state
\begin{equation}
Z_{m}^{n,t+1}={F_{\text{RBIG}}^{t}}^{-1}\left({Z_{m}^{n,t+1}}^\text{RBIG}\right),\label{6}
\end{equation}
where ${F_{\text{RBIG}}^{t}}^{-1}$ is an inverse RBIG transformation and $Z_{m}^{n,t+1}$ are back-transformed updated neighbourhood realisations.

\item Replacing the previous block model realisations with updated results.

\end{enumerate}

\subsection{Proposed joint updating of grades and geological domains}\label{subsec2.3}

We propose a rapid updating algorithm inspired by the applications of PGS in the cascade modelling of geological domains and their corresponding grades. To achieve this, EnKF-MDA is first used to update simulated realisations of GRFs with new observations, and then to separately update grades within the updated domains. Figure~\ref{fig1} shows the complete flowchart of the proposed approach for the joint rapid updating of multivariate grades and geological domains. The steps of the proposed algorithm are as follows:

\begin{figure}[ht]%
\centering
\includegraphics[width=1\textwidth]{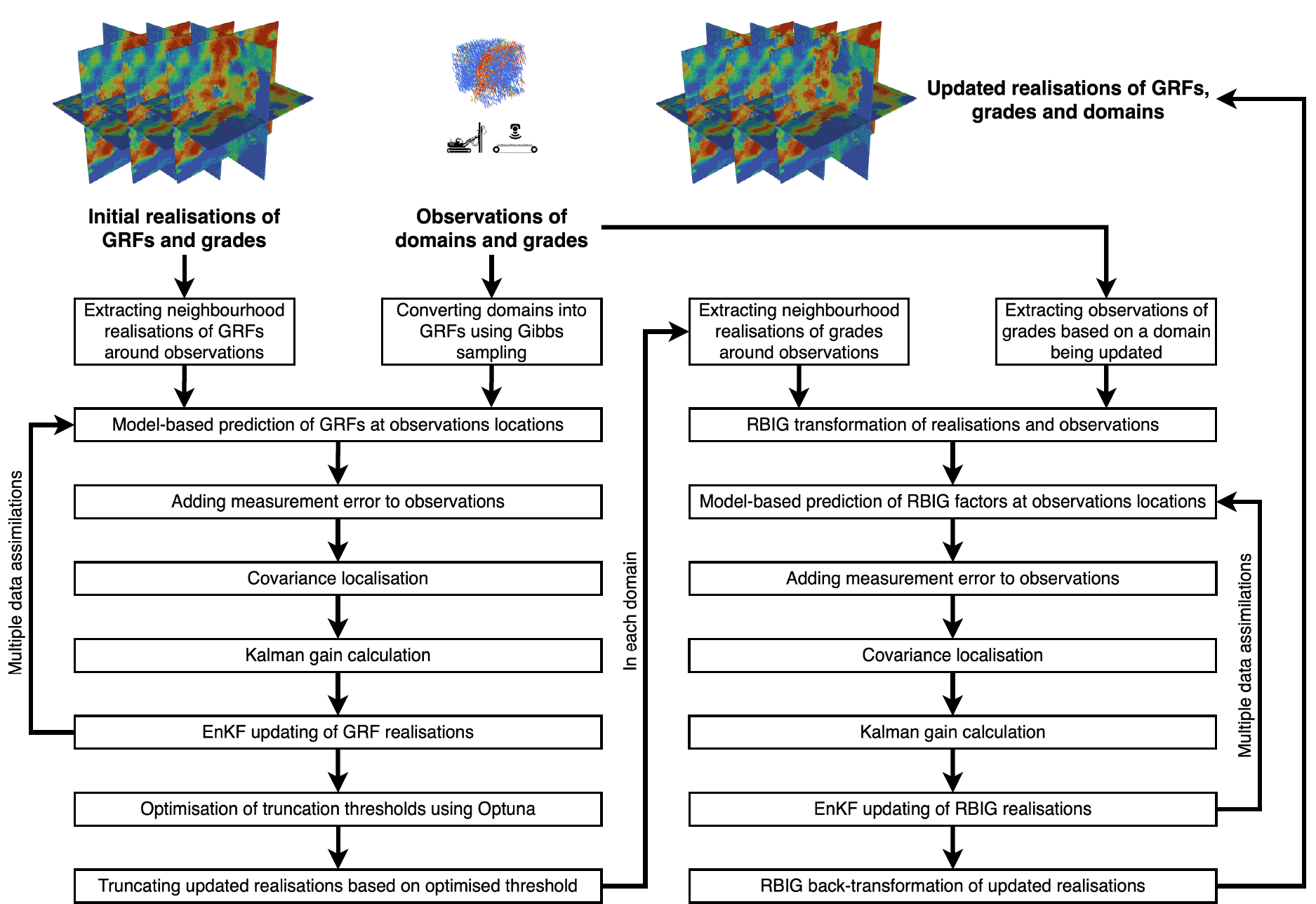}
\caption{The proposed approach for joint rapid updating of grades and geological domains.}\label{fig1}
\end{figure}

\begin{enumerate}[1.]

\item Extracting the neighbourhood around new observations from the prior block model realisations. In general, the neighbourhood within 2-3 blocks away from the observations offers a good balance between computation time and targeting the most important blocks for short-term decisions.

\item Applying the Gibbs sampler to new observations using the same truncation rule and the variograms that were used during the prior PGS modelling of geological domains. It is recommended that the number of iterations for the Gibbs sampler remains the same as in the prior modelling.

\item Applying EnKF-MDA to update prior realisations of GRFs based on observations of domains converted to GRFs in step 2.

\item Deriving optimal truncation thresholds using the hyperparameter tuning library Optuna \citep{bib45} by cross-validating the results against all available observations.

\item Truncating the updated realisations based on the optimised truncation thresholds.

\item Replacing the previous GRF and domain realisations with updated models.

\item In each domain, separately updating cross-correlated grade variables using the EnKF-MDA + RBIG approach described in Section~\ref{subsec2.2}.

\end{enumerate}

\section{Results}\label{sec3}

\subsection{Overview of a case study}\label{subsec3.1}

The proposed algorithm is suitable for any deposit type, as it employs methods that have previously produced good results across a range of deposits. This paper showcases the performance of pluri-Gaussian rapid updating using both a synthetic and a real case study based on data from an IOCG deposit located in South Australia. The original drill hole data comprises nearly 40,000 samples of five geological domains and three cross-correlated grade variables (Au, Cu, and U). The domains are volcanic (VOLC), hematite (HEM), dolomite (DOLM), dolerite (DOLT) and skarn (SKRN). Copper and gold mineralisation primarily occurs in hematite breccias, whereas uranium is regarded as a deleterious element. Table~\ref{tab1} presents the key statistical parameters for Au, Cu, and U grades. It is important to note that all three grade variables exhibit highly skewed distributions.

\begin{table}[ht]
\begin{center}
\caption{Statistical parameters of the grade variables in drill hole samples.}\label{tab1}%
\begin{adjustbox}{width=0.7\textwidth}
\begin{tabular}{lccc}
\toprule
Parameter & Au (ppm) & Cu (\%) & U (ppm)\\
\midrule
Mean & 0.29 & 0.50 & 50.80\\
Standard deviation & 0.51 & 1.04 & 77.16\\
Skewness & 5.34 & 4.17 & 6.87\\
Kurtosis & 46.15 & 23.64 & 87.61\\
\botrule
\end{tabular}
\end{adjustbox}
\end{center}
\end{table}

\begin{figure}[p]%
\centering
\includegraphics[width=1\textwidth]{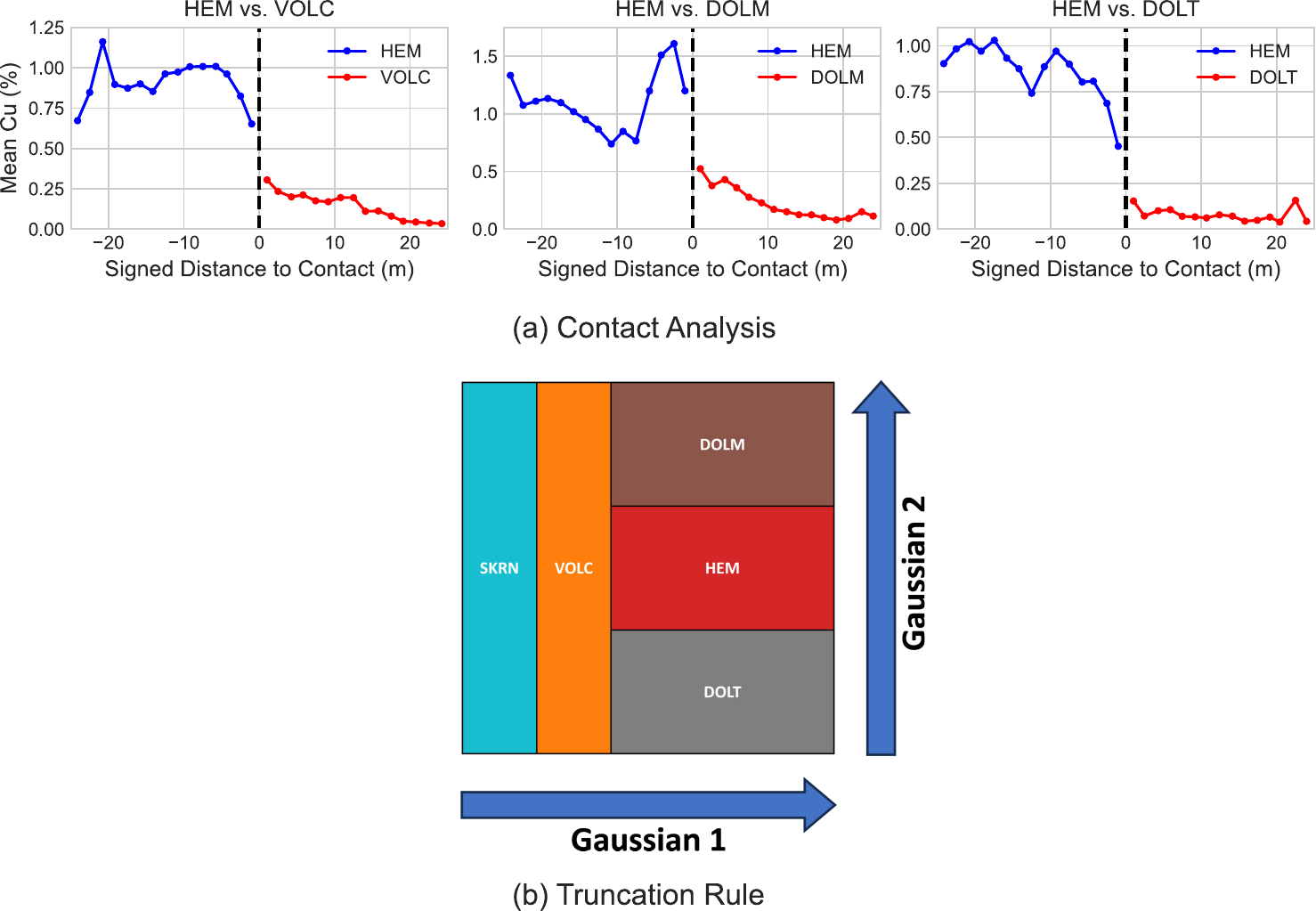}
\caption{(a) Contact plots between HEM and neighbouring domains, and (b) truncation rule.}\label{fig2}
\end{figure}

Table~\ref{tab2} compares geological domains based on global proportions and grade distributions. HEM has the highest proportion and the highest average grades among the various domains. DOLM has a similar proportion to HEM but with significantly lower and less deviated grades. VOLC has slightly lower gold and copper grades than DOLM, but a similar average uranium grade. In contrast, DOLT and SKRN are much smaller in proportion and exhibit very low grades. Figure~\ref{fig2} illustrates the contact analysis between HEM and its neighbouring domains, VOLC, DOLM and DOLT, as well as the corresponding truncation rule for PGS. Notably, the transition of mean copper grade from HEM to other domains can be described as a hard boundary. In this study, 100 prior realisations of geological domains were modelled using PGS with 1,000 Gibbs sampler iterations. The block model dimensions are 5 m × 5 m × 5 m, with 101 × 81 × 101 blocks.

\begin{table}[ht]
\begin{center}
\caption{Statistical parameters of the geological domains, including the mean (M) and standard deviation (SD) of grades for each domain.}\label{tab2}%
\begin{adjustbox}{width=1\textwidth}
\begin{tabular}{lccccccc}
\toprule
Domain & Proportion (\%) & M Au & M Cu & M U & SD Au & SD Cu & SD U\\
\midrule
VOLC & 14.20 & 0.14 & 0.19 & 29.92 & 0.32 & 0.43 & 41.83\\
HEM & 40.04 & 0.52 & 0.93 & 85.39 & 0.58 & 1.33 & 100.75\\
DOLM & 36.48 & 0.17 & 0.26 & 30.19 & 0.47 & 0.76 & 44.86\\
DOLT & 6.92 & 0.03 & 0.08 & 15.25 & 0.10 & 0.16 & 34.04\\
SKRN & 2.33 & 0.02 & 0.11 & 12.43 & 0.04 & 0.15 & 19.71\\
\botrule
\end{tabular}
\end{adjustbox}
\end{center}
\end{table}

\begin{figure}[p]%
\centering
\includegraphics[width=0.65\textwidth]{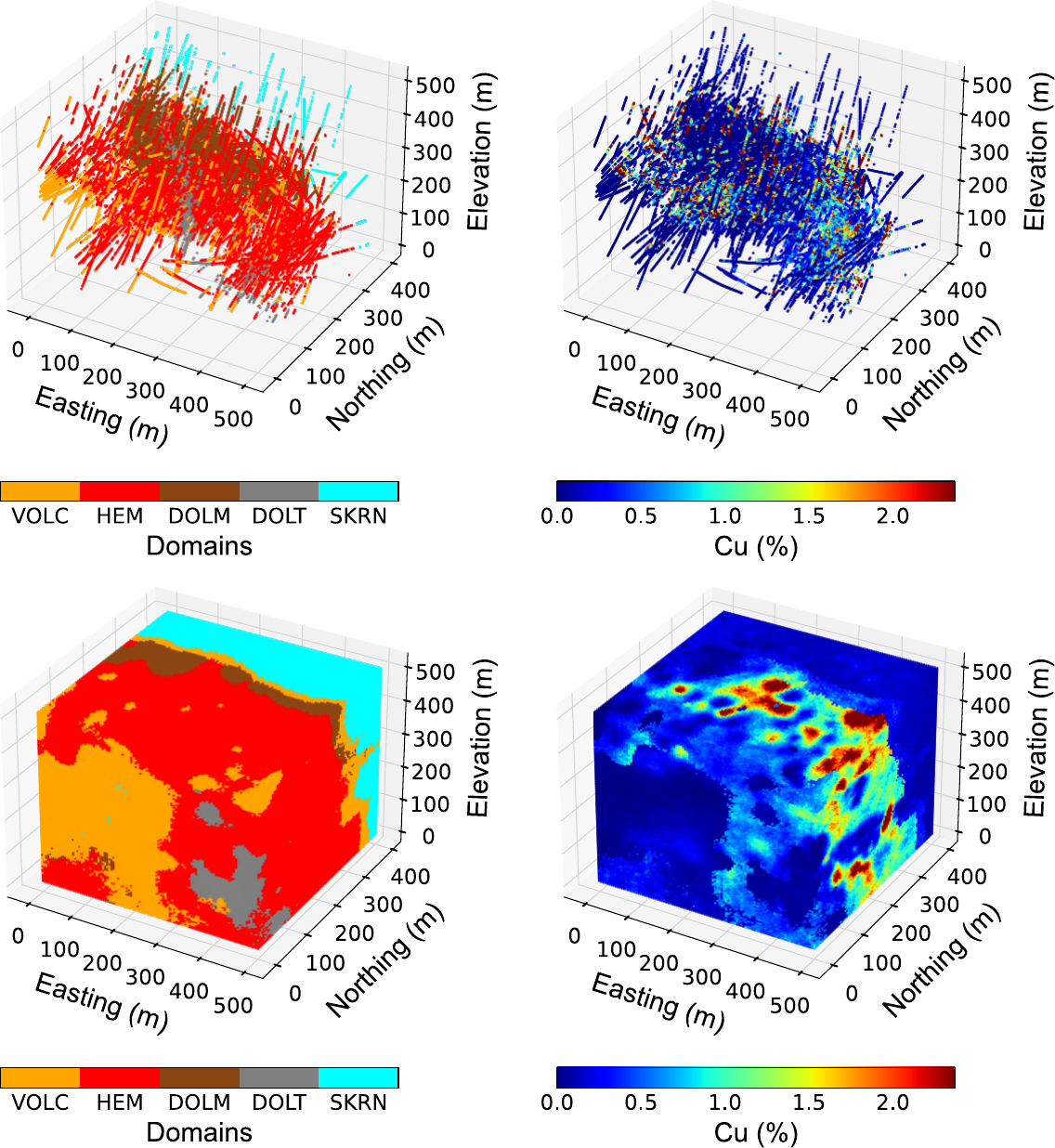}
\caption{3D view of geological domains and copper grade in drill hole samples (top) and prior block model. The most probable domains from 100 PGS realisations and the e-type copper model are presented for the prior block model.}\label{fig3}
\end{figure}

\begin{figure}[ht]%
\centering
\includegraphics[width=0.77\textwidth]{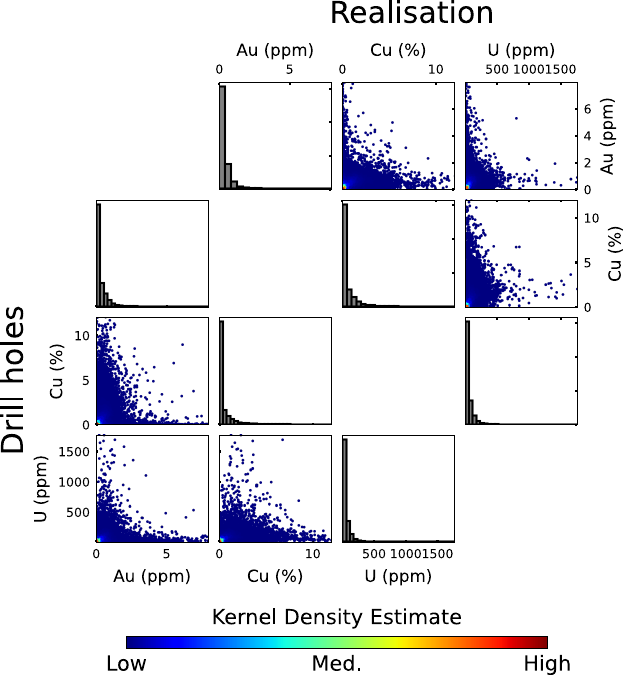}
\caption{Scatter plots between grade variables alongside their respective histograms for drill hole samples and one block model realisation.}\label{fig4}
\end{figure}

In each domain, RBIG transformed cross-correlated grade variables into multi-Gaussian factors. The turning bands simulation was then used to compute the 100 RBIG realisations of Au, Cu, and U, followed by back-transformation to the original space. Figure~\ref{fig3} illustrates a 3D view of the drill hole samples and a prior block model. As copper is highly skewed, the colour bar is adjusted to the 95th percentile to better represent the high-grade zones. In the prior block model, the most probable categories are displayed for the domain model and the e-type for the copper model. The comparison of univariate and multivariate distributions shows that RBIG helped to reproduce the highly skewed histograms of Au, Cu, and U, while also honouring the multivariate relationships (Figure~\ref{fig4}).

\subsection{Synthetic case study}\label{subsec3.2}

\begin{figure}[ht]%
\centering
\includegraphics[width=1\textwidth]{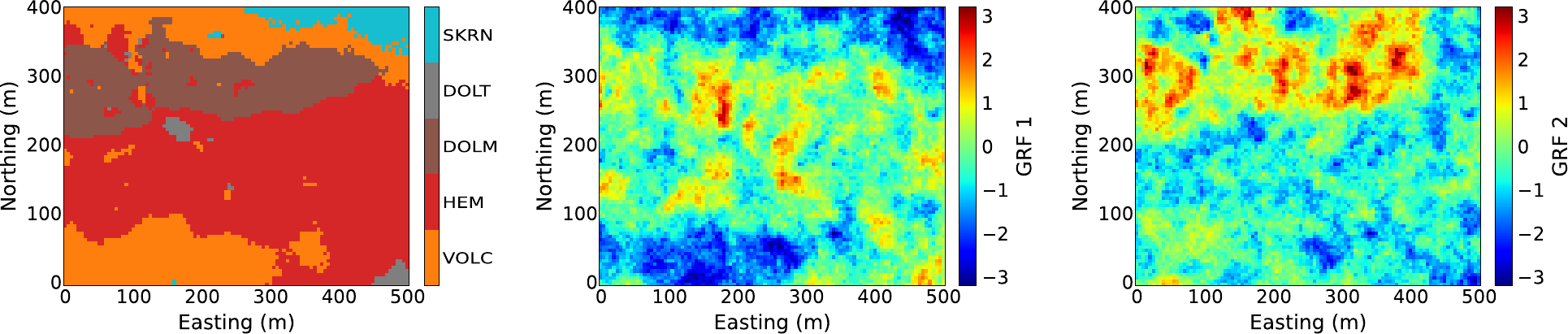}
\caption{From left to right: the most probable domains from prior realisations, a single realisation of the first and second GRFs.}\label{fig5}
\end{figure}

\begin{figure}[ht]%
\centering
\includegraphics[width=1\textwidth]{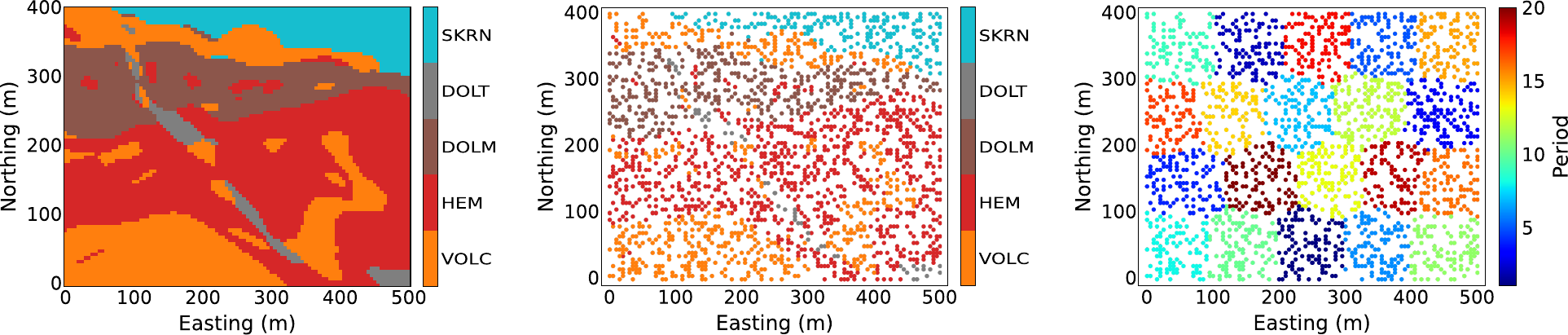}
\caption{From left to right: the ground truth, observations of geological domains, and time periods for sequential updating.}\label{fig6}
\end{figure}

In this synthetic case study, we used a 2D section of the prior block model at an elevation of 350 m. Figure~\ref{fig5} illustrates the most probable domains obtained from the 100 PGS realisations and two GRFs in the first realisation. The outcome of this synthetic experiment relies on the ground truth model, which represents a true but unknown environment. Therefore, it is crucial to model the ground truth using a different methodology from PGS to avoid introducing unnecessary biases. DomainMCF geological modelling software \citep{bib41} was applied to create a ground truth model based on the same drill hole data as the PGS model. A total of 2045 observations (about 25\% of the ground truth) were randomly sampled and subdivided into 20 time periods to use for sequential rapid updating (Figure~\ref{fig6}). It should be noted that the produced ground truth model may be slightly inconsistent with the truncation rule used for the prior modelling. However, such inconsistencies are important in ensuring this synthetic case study remains fair and unbiased, since similar issues can also occur in real cases.

\begin{figure}[ht]%
\centering
\includegraphics[width=1\textwidth]{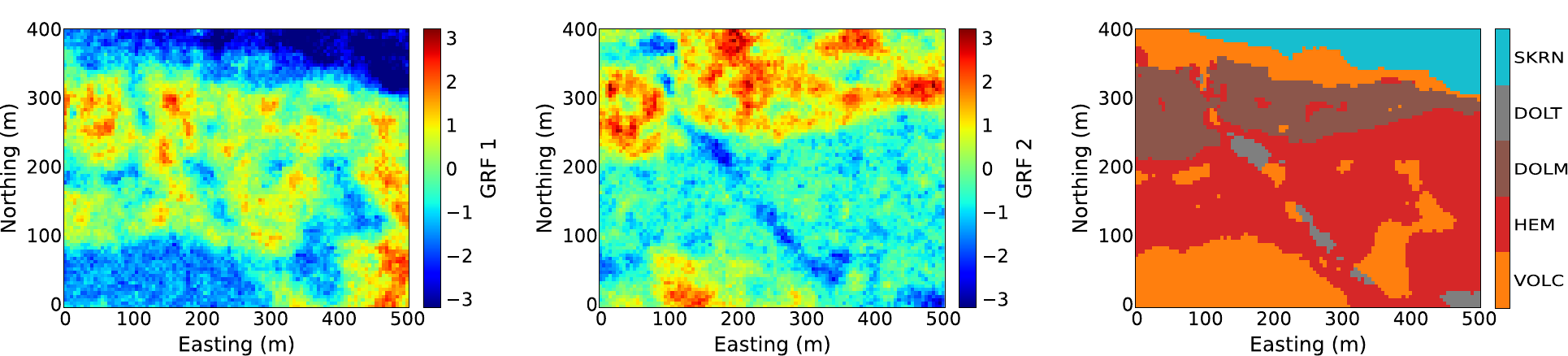}
\caption{From left to right: a single realisation of the first and second GRFs and the most probable domains from 100 updated realisations.}\label{fig7}
\end{figure}

At each time period, geological domains were converted into two GRFs using the Gibbs sampler, based on the prior variograms and the truncation rule. For simplicity, a random Gaussian error with a standard deviation of 0.1 was added to the GRF observations to introduce a measure of geological logging uncertainty. Then, the EnKF-MDA updated the prior GRF realisations based on new observations in 20 time periods and truncated them back to the categorical space. For EnKF-MDA, the number of data assimilations depends on user preference, as more assimilations improve performance but also take more time. Typically, 4-5 data assimilations offer a good balance between performance and computation time. Similar to \cite{bib26}, this paper performed 10 data assimilations for EnKF-MDA to prioritise better performance. Figure~\ref{fig7} shows the final updated GRFs in the first realisation and the most probable model, which closely matches the ground truth. It is also visually noticeable how the updating improved the reproduction of the long-range structure of the DOLT domain, even on the second GRF. Another difference is the SKRN domain on the top part of the model, which is now much larger.

The updated model achieves a 92.89\% accuracy in matching the ground truth. However, its accuracy in the DOLT domain is only around 67\%, while in other domains it varies from 89\% to 96\%. Table~\ref{tab3} shows that DOLT accuracy can be significantly enhanced by increasing the truncation threshold between DOLT and HEM. Conversely, improving the DOLT accuracy leads to a decrease in HEM accuracy, which accounts for a larger proportion and higher grades. Therefore, it is important to carefully calibrate the truncation thresholds to ensure both overall and individual accuracies remain high.

\begin{table}[ht]
\begin{center}
\caption{HEM and DOLT accuracies after increasing the corresponding truncation threshold.}\label{tab3}%
\begin{adjustbox}{width=0.8\textwidth}
\begin{tabular}{lccccccc}
\toprule
\multirow{2}{*}{Domain} & \multicolumn{7}{c}{Accuracies after increasing the truncation threshold by} \\
& +0\% & +5\% & +10\% & +15\% & +20\% & +25\% & +30\%\\
\midrule
HEM & 96.36 & 96.21 & 95.77 & 94.72 & 93.57 & 91.40 & 88.29\\
DOLT & 66.93 & 73.93 & 80.54 & 82.10 & 84.44 & 86.77 & 88.72\\
\botrule
\end{tabular}
\end{adjustbox}
\end{center}
\end{table}

The hyperparameter tuning library Optuna \citep{bib45} was used to compute optimal truncation thresholds through cross-validation with all available observations. In this case, the standard accuracy is not an effective metric given the imbalanced data. Instead, we consider the combination of the F1 score and the geometric mean (G-Mean). The F1 score provides a balanced measure of precision and recall by considering both false positives and false negatives. The G-Mean is the geometric mean of the per-class recall, making it highly sensitive to poor performance in any single class. This combination ensures that the model performs well overall without completely failing in small but important domains. The objective function for Optuna in this case study is as follows:

\begin{equation}
\text{Score}=w_1\text{F1}+w_2\text{G-Mean},\label{7}
\end{equation}

where the weights were set equally for this study ($w_1=0.5,\ w_2=0.5$), but can be adjusted depending on whether the emphasis is on achieving overall predictive power or on ensuring robust performance in less prevalent domains.

\begin{figure}[ht]%
\centering
\includegraphics[width=1\textwidth]{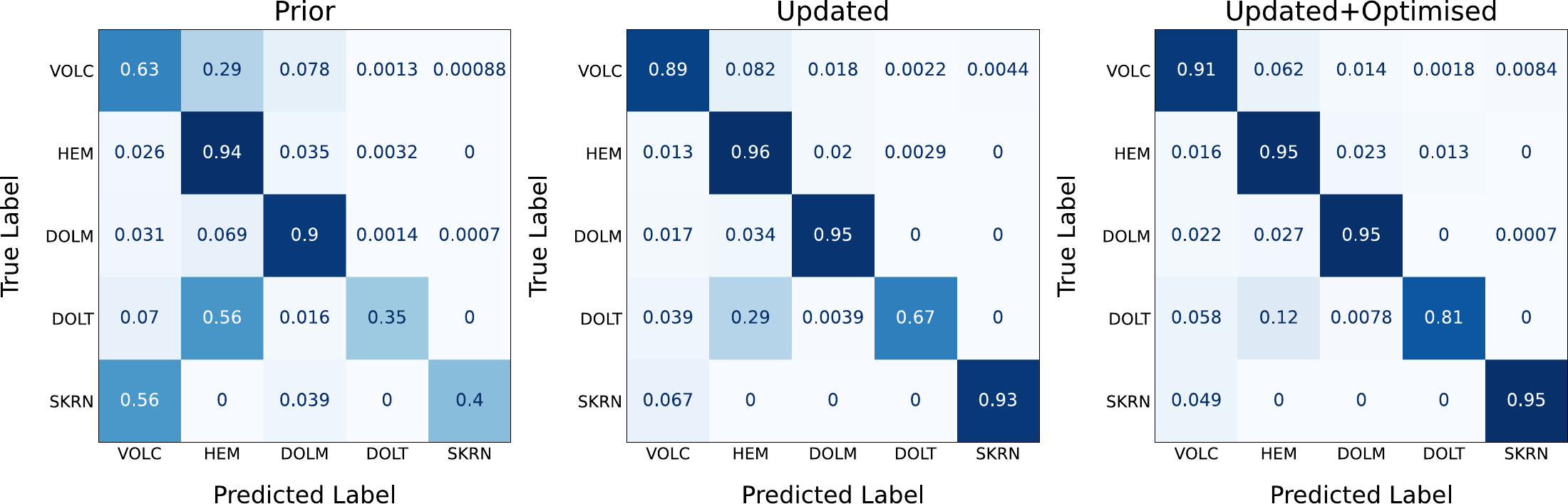}
\caption{Confusion matrices between true and predicted domain labels for the prior and updated models, and the updated model with optimal thresholds.}\label{fig8}
\end{figure}

The confusion matrices in Figure~\ref{fig8} clearly illustrate the difference between the prior and updated models. The prior model, which matched only 77\% of the ground truth, frequently confused SKRN with VOLC, DOLT with HEM, and at times VOLC with HEM. Updating the realisations improved the accuracy of all the domains except DOLT. Finally, the optimal thresholds increased overall accuracy to 93.46\%, significantly improved DOLT domain prediction and provided slightly better results for VOLC and SKRN.

\begin{figure}[ht]%
\centering
\includegraphics[width=0.8\textwidth]{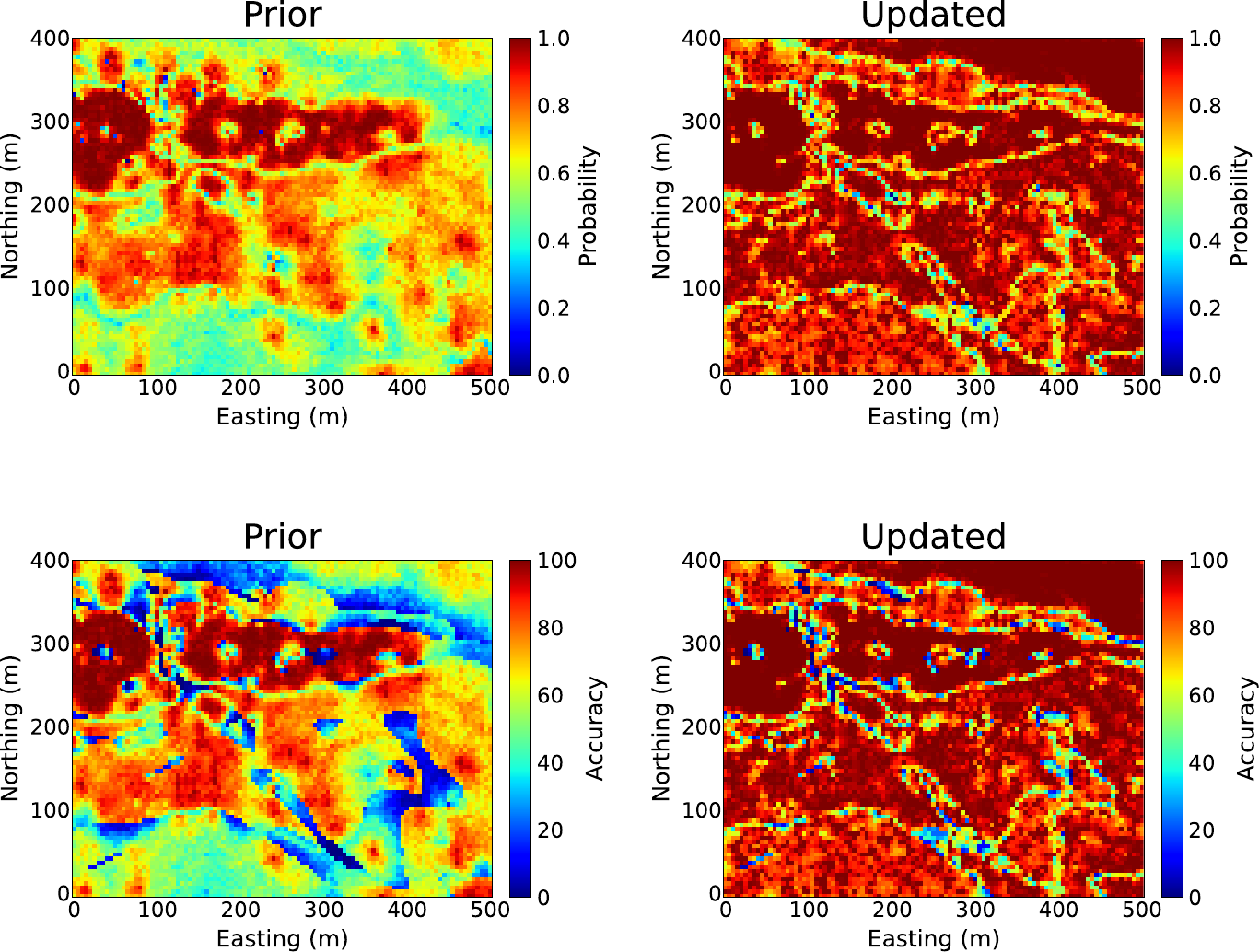}
\caption{Probability of the most probable domain (top) and accuracy compared to the ground truth (bottom) for prior and updated models.}\label{fig9}
\end{figure}

A more detailed comparison of the prior and updated models is shown in Figure~\ref{fig9}. The probabilities of the most probable domains indicate that the biggest challenge for PGS was the uncertainty due to insufficient conditioning data, which resulted in lower probabilities for VOLC, DOLT, and SKRN. The rapid updating significantly minimised uncertainty for the entire model, apart from the boundaries between domains. However, this is realistic, given that observations are also slightly uncertain, which can affect truncation and sometimes cause mismatches between neighbouring domains.

The accuracy results, defined by the rate of matching between realisations and the ground truth, present a more interesting picture. In the prior model, significant inaccuracies occurred in the regions with the lowest probabilities and in the long-range structures of DOLT and VOLC. As expected, the prior predictions of HEM and DOLM were mostly accurate, except for the contacts with other domains. After the updates, mismatches are primarily located at domain boundaries, which can be attributed to the complexity of contact relationships. The proposed algorithm is currently limited to two GRFs, which may not be sufficient to fully define all contact relationships.

\subsection{Real case study}\label{subsec3.3}

\begin{figure}[ht]%
\centering
\includegraphics[width=1\textwidth]{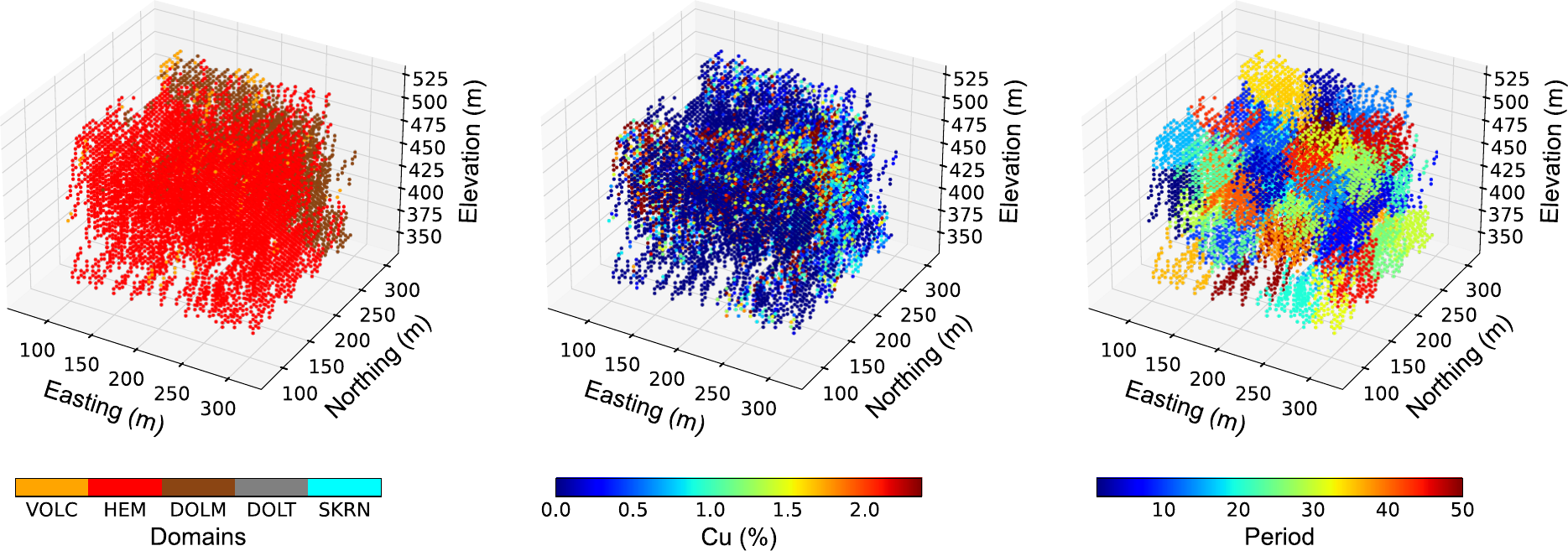}
\caption{3D view of geological domains, copper grade and time periods in RC observations.}\label{fig10}
\end{figure}

In the real case study, approximately 50,000 RC drilling samples, taken at 1m intervals, were available for use as observations for rapid updating. For simplicity, these samples were upscaled to match the resource model dimensions, resulting in 9,712 observations. There is a total of 7,180 HEM, 2,248 DOLM, and 283 VOLC observations, along with the Au, Cu, and U grades. Figure~\ref{fig10} illustrates a 3D view of observations of geological domains, copper grades, and 50 time periods at which rapid updating will proceed. The statistics of the observations presented in Table~\ref{tab4} are similar to the global statistics for gold and uranium from the original drill hole samples. However, the observations exhibit slightly higher and less skewed copper grades.

\begin{table}[ht]
\begin{center}
\caption{Statistical parameters of the grade variables in observations.}\label{tab4}%
\begin{adjustbox}{width=0.7\textwidth}
\begin{tabular}{lccc}
\toprule
Parameter & Au (ppm) & Cu (\%) & U (ppm)\\
\midrule
Mean & 0.33 & 0.76 & 52.62\\
Standard deviation & 1.26 & 1.04 & 88.12\\
Skewness & 4.16 & 2.64 & 6.49\\
Kurtosis & 38.89 & 9.02 & 64.86\\
\botrule
\end{tabular}
\end{adjustbox}
\end{center}
\end{table}

\begin{figure}[ht]%
\centering
\includegraphics[width=0.7\textwidth]{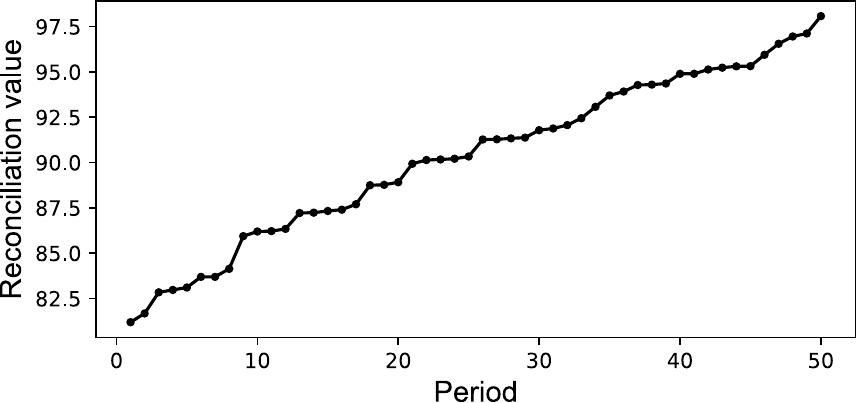}
\caption{Reconciliation between model-based predictions and all observations at different time periods.}\label{fig11}
\end{figure}

Due to the unavailability of the actual dates of the RC drilling, observations were subdivided into 50 time periods based on the clustering of coordinates. At each time period, the updating process begins by extracting the neighbourhood around new observations. It is generally recommended for the neighbourhood to be three blocks away from observations to target blocks that are important for short-term decisions and to ensure that updates can be done in near-real time. Geological domain observations are then converted into two GRFs based on the truncation rule and variograms used during the prior PGS modelling. Next, both prior realisations of GRFs and new observations served as inputs for EnKF-MDA to perform rapid updating. It is important to note that observations do not include the SKRN and DOLT domains, and the number of VOLC samples is relatively small. Figure~\ref{fig11} illustrates the reconciliation between the most probable predictions at observation locations and the entire observation data across all time steps. After 50 updates, the proposed pluri-Gaussian rapid updating increases the accuracy from 81\% to nearly 98\%.

\begin{figure}[ht]%
\centering
\includegraphics[width=0.65\textwidth]{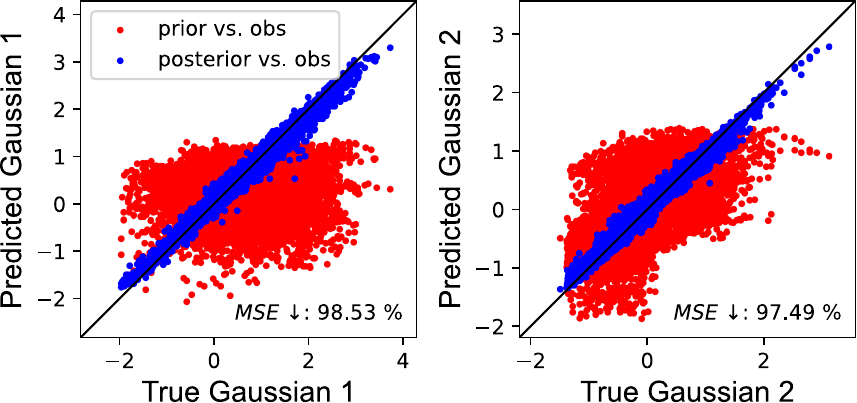}
\caption{Predictions versus observations plots before and after rapid updating for two GRFs.}\label{fig12}
\end{figure}

\begin{figure}[ht]%
\centering
\includegraphics[width=1\textwidth]{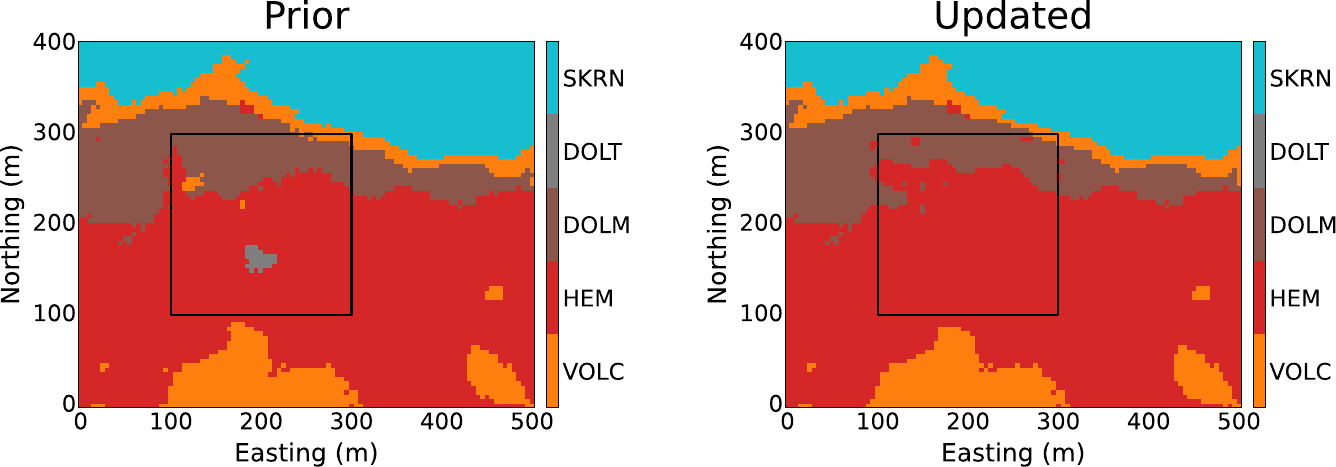}
\caption{2D view of prior and updated most probable domains at 390 m elevation.}\label{fig13}
\end{figure}

As noted in the synthetic case study, the truncation thresholds are sensitive to uncertainties, particularly those associated with hard domain boundaries. The primary purpose of rapid updating is to integrate real-time observations into resource models, thereby facilitating rapid decision-making under uncertainty. Therefore, the uncertainty in domain boundaries will remain a major challenge, especially for the proposed methodology, which is highly dependent on contact relationships. Even though the final updated GRF realisations closely match the observations (see Figure~\ref{fig12}), mismatches can still occur between neighbouring domains. Overall, the results show that the updated domains accurately match the observations, with 98.11\% accuracy and reductions in mean squared error (MSE) of 98.53\% and 97.49\% for the two GRFs. Figure~\ref{fig13} illustrates the prior and updated most probable geological domains at an elevation of 390 m. The significant difference is that rapid updating removed the DOLT domain and a small VOLC structure. This may have an important impact on the economic value of this section of the deposit, as both of these domains have lower average grades.

The next step is to update the cross-correlated Au, Cu and U grades within each domain. For this, RBIG is a recommended addition to EnKF-MDA, as it can quickly transform multivariate data into independent multi-Gaussian factors and ensure the reproduction of multivariate relationships \citep{bib34,bib26}. However, RBIG and PPMT struggle to reproduce extreme values in highly skewed cases. This can also be true for the conventional geostatistical co-simulation methods. One way to minimise deviations between extreme values of predictions and observations is to use FA \citep{bib40,bib21}, as it produces fewer artifacts in the presence of extreme values than RBIG and PPMT \citep{bib35}. Nevertheless, FA is currently considerably slower than PPMT, which, in turn, is slower than RBIG. Another recent solution is a non-monotonic transformation of GRFs, which has demonstrated better reproduction of the tails of skewed distributions compared to PPMT and co-simulation \citep{bib42}. This method is, unfortunately, too complex and time-consuming, as it involves multiple indicator variables, three stages of model parameter inference, simulation, and particle filtering.

\begin{figure}[ht]%
\centering
\includegraphics[width=1\textwidth]{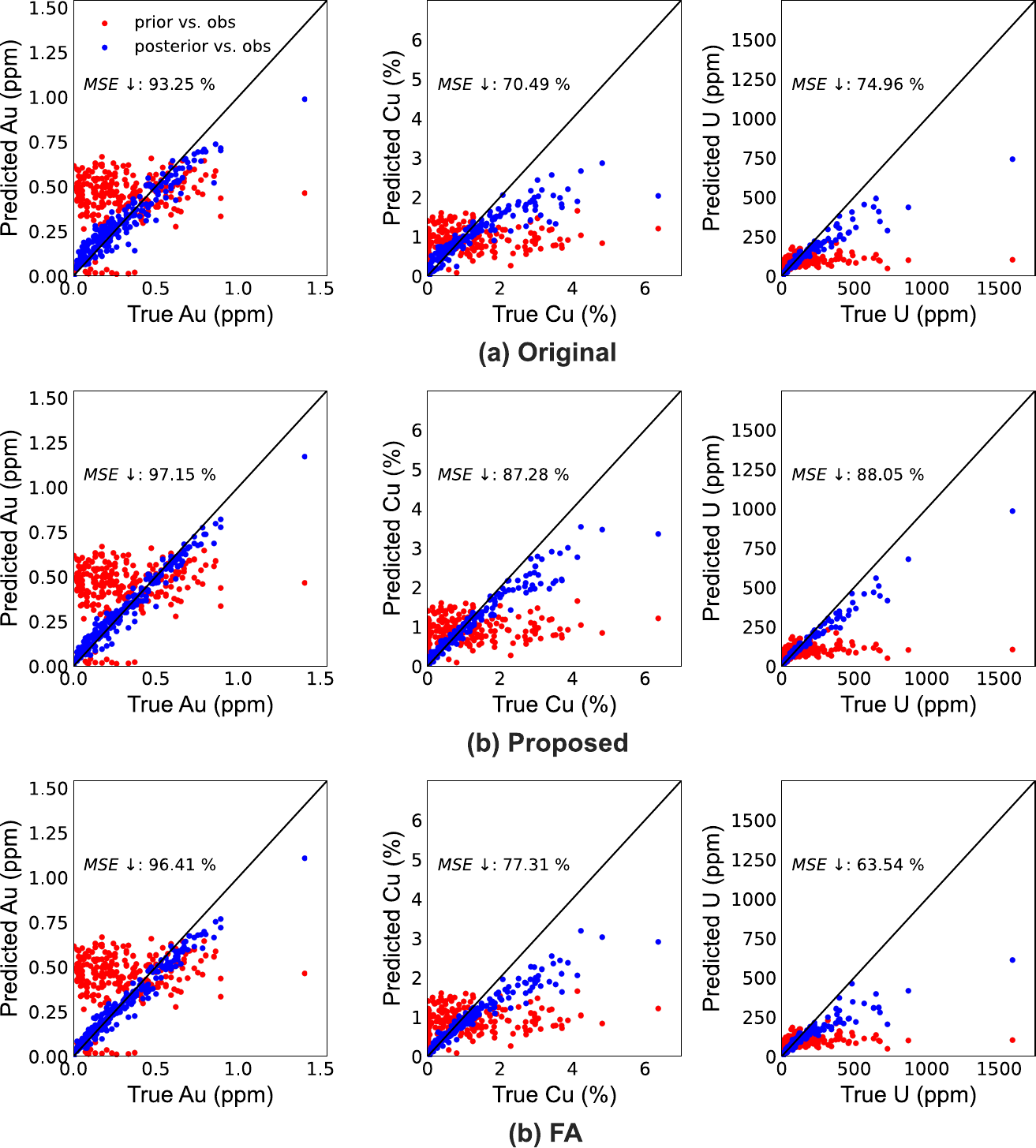}
\caption{Predictions versus observations plots before and after the rapid updating of Au, Cu, and U using the original and proposed RBIG approaches, as well as FA for the period 10.}\label{fig14}
\end{figure}

\begin{figure}[ht]%
\centering
\includegraphics[width=1\textwidth]{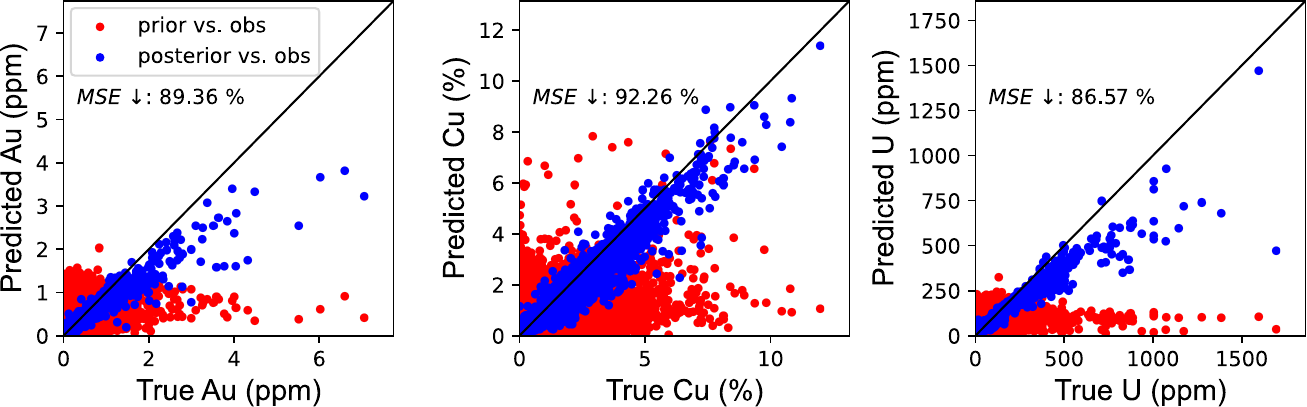}
\caption{Final predictions versus observations plots before and after the rapid updating of Au, Cu and U after 50 periods.}\label{fig15}
\end{figure}

This study presents a simple approach to separately update extreme values, addressing the issue of reproducing skewed distributions. As all three variables are heavily right-skewed to a similar extent, the 95th percentile is selected as the threshold for extreme values. Figure~\ref{fig14} shows plots of predictions against observations before and after the rapid updating in period 10 for both the original and proposed RBIG approaches, with the FA included for comparison. Visually, clear extreme values are evident, especially in the prior copper and uranium predictions, which deviate from observations and result in only 70-75\% MSE reductions with the original RBIG. The proposed method improved results for all three variables, particularly copper and uranium. Overall, the improvements compared to the original are 3.9\% for gold, 16.79\% for copper, and 13.09\% for uranium. FA only improved error reductions for gold and copper but performed worse for uranium compared to the original RBIG approach. The final results after updating 50 periods of observations demonstrate that the proposed approach performs well across all three variables, with 87-92\% reductions in error (Figure~\ref{fig15}).

\begin{figure}[ht]%
\centering
\includegraphics[width=1\textwidth]{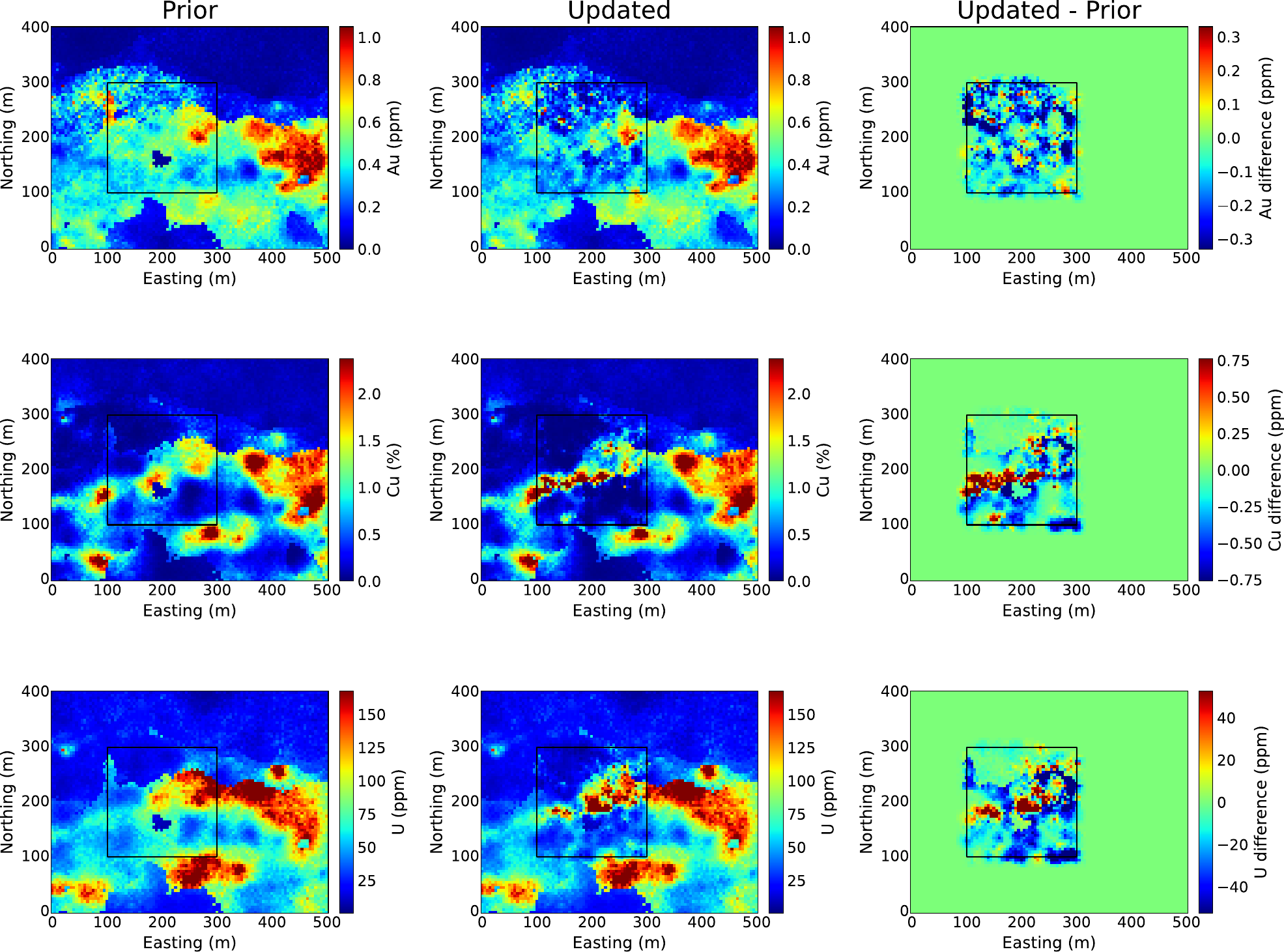}
\caption{2D view of prior and updated Au, Cu and U e-type models at 390 m elevation, as well as the difference between them. The colour bars are adjusted to better visualise the high-grade zones and differences in grades.}\label{fig16}
\end{figure}

Figure~\ref{fig16} illustrates the average grades of gold, copper, and uranium in a 2D section at an elevation of 390 m before and after rapid updating, along with the difference between the updated and prior models. The colour bars were adjusted to better visualise the high-grade zones and the difference between grades. The orebody with high copper content is significantly narrower and more connected in the updated model, and the same can be said for the high-uranium zone. This factor will be important to consider when calculating the economic value of these blocks since uranium is a deleterious element in this deposit. Overall, the rapid updating helped to minimise the overestimation of grades and corrected the results. In a practical scenario, near-real-time joint updating of geological domains and grade variables would enable rapid decision-making, including mine planning and optimisation of comminution and processing.

The key difference is the removal of the DOLT domain, as shown in Figure~\ref{fig13}. The DOLT appears visually out of place when examining models of prior grades, appearing as a low-grade structure within a moderate-to-high grade zone. In contrast, the updated model removed the DOLT and provided a more realistic grade distribution. Table~\ref{tab5} presents statistical metrics comparing the prior and updated models for each grade variable at observation locations, which were mislabelled in the prior model as DOLT. For instance, the mean grades in the updated model are now closer to the truth. The updated model also achieves much higher R2 scores and reduces MSE by 79-87\% compared to the prior model.

\begin{table}[ht]
\begin{center}
\caption{Statistical parameters of the grade variables at observation locations that were originally labelled as DOLT.}\label{tab5}%
\begin{adjustbox}{width=0.7\textwidth}
\begin{tabular}{lccc}
\toprule
Parameter & Au (ppm) & Cu (\%) & U (ppm)\\
\midrule
True mean & 0.24 & 0.33 & 48.00\\
Prior mean & 0.03 & 0.14 & 21.67\\
Updated mean & 0.18 & 0.22 & 36.76\\
Prior $R^2$ & 0.10 & 0.01 & 0.00\\
Updated $R^2$ & 0.91 & 0.90 & 0.99\\
MSE reduction & 87.40\% & 83.86\% & 79.28\%\\
\botrule
\end{tabular}
\end{adjustbox}
\end{center}
\end{table}

\section{Conclusions}\label{sec5}

This paper presents a data assimilation algorithm for jointly updating geological domains and cross-correlated grade variables. The concept is inspired by the PGS algorithm, a widely used geostatistical simulation method for modelling facies and domains. The Gibbs sampler provides a straightforward way for converting categorical data into GRFs, which are then suitable for updating with EnKF-MDA. Following the principles of cascade modelling, the next step is to update cross-correlated grade variables within each geological domain. Overall, the proposed approach has demonstrated its effectiveness in both synthetic and real case studies based on data from an IOCG deposit in South Australia.

The proposed methodology builds upon the previous approach presented in \cite{bib26}. First, this study introduces an algorithm to jointly update cross-correlated grades and geological domains, rather than focusing solely on grades. The PGS approach was included to accurately update domains and account for complex contact relationships. Additional optimisation of truncation thresholds provides greater flexibility for adapting to new data. Furthermore, the approach addresses the challenge of accurately updating highly skewed distributions, a limitation observed with the previous combined use of EnKF-MDA and RBIG. By separately updating extreme values, we significantly reduced errors across all three grade variables compared to the original use of RBIG and FA.

However, some limitations still need to be addressed in future research. Firstly, the proposed approach assumes that prior realisations were modelled using PGS and that GRF realisations are always available. This assumption was made to simplify the method and reduce computation time, as converting hundreds of block model realisations into GRFs using the Gibbs sampler for each time period is computationally intensive. Moreover, the algorithm is currently limited to two GRFs, which may not be sufficient for more complex scenarios. Finally, to simplify the case studies, this paper does not address the change of support problem. In the real case study, RC drilling samples were upscaled to match the resource model dimensions by averaging their grades over a block volume. However, in practice, observations may come from different sources, and a rapid updating algorithm should account for the change of support.

The proposed method follows the principles of cascade modelling by modelling geology first and grades later, which is effective for hard domain boundaries. However, for soft boundaries, the joint simulation approach is usually recommended \citep{bib46}. This method typically requires joint variography between normal-score-transformed grade variables and GRFs derived from PGS modelling. Future research will investigate the applicability of multi-Gaussian transforms as an alternative to joint variography.

To make the algorithm truly comprehensive and capable of updating the entire resource knowledge, it must also incorporate geometallurgical information. However, since geometallurgical variables related to comminution and processing are often non-additive, their geostatistical modelling and rapid updating with EnKF may not be the most appropriate strategy. In future research, we will focus on extending the current algorithm to integrate non-additive data. For example, rapid resource model updating can be incorporated into the primary-response framework \citep{bib43}, as updating primary variables enhances the prediction of response parameters.

\backmatter

\bmhead{Acknowledgments} 

The research reported here was supported by the Australian Research Council Industrial Transformation Training Centre for Integrated Operations for Complex Resources (ARC ITTC IOCR - project number IC190100017) and funded by universities, industry and the Australian Government.

\bibliography{bibliography}

\end{document}